\documentstyle[12pt,epsf]{article}
\renewcommand{\bar}[1]{\overline{#1}}

\newcommand{\mathrm}{{\rm}} 

\begin{document}

\begin{flushright} 
hep-ph/9810517 \\CPT-98/P.3706 \\BIHEP-TH-98-8 
\end{flushright}

\bigskip\bigskip
\centerline{\large \bf Quark Flavor Separation in 
$\Lambda$-Baryon
Fragmentation}

\vspace{22pt}

\centerline{\bf Bo-Qiang Ma$^{a}$ and Jacques Soffer$^{b}$}

\vspace{8pt}
{\centerline {$^{a}$CCAST (World Laboratory), P.O.~Box 8730, Beijing 
100080, China}}

{\centerline {and Institute of High Energy Physics, Academia Sinica,
P.~O.~Box 918(4),} 

{\centerline {Beijing 100039, China\footnote{Mailing address}} 

{\centerline{e-mail: mabq@hptc5.ihep.ac.cn}} 

{\centerline {$^{b}$Centre de Physique Th\'eorique, 
CNRS, Luminy Case 907,}} 

{\centerline  {F-13288 Marseille Cedex 9, France}}

{\centerline{e-mail: soffer@cpt.univ-mrs.fr}}


\vspace{10pt}
\begin{center} {\large \bf Abstract}

\end{center}

It is shown that 
neutrino and antineutrino
deep inelastic scattering 
of unpolarized and polarized $\Lambda$
and $\bar{\Lambda}$ productions, can provide a
clean separation of unpolarized and polarized 
fragmentation functions 
of a quark into a $\Lambda$, 
for both light-flavor quarks and antiquarks and also for strange
quarks. Combining with $\Lambda$ and $\bar{\Lambda}$ productions in
polarized electron deep inelastic scattering, 
one can systematically measure or check the various flavor and spin
dependent fragmentation functions.
 Such measurements can provide crucial tests of different
predictions concerning the spin structure of hadrons
and the quark-antiquark asymmetry of the nucleon
sea.

\vfill
\centerline{PACS numbers: 13.87.Fh, 13.15.+g, 13.60.Hb, 14.20.Jn} 
\vfill

\centerline{Accepted for publication in Physics~Review~Letters} 
\newpage


Spin physics is currently one of the most active research directions
of hadron physics due to the rich phenomena, which are
different from
the naive theoretical expectations and the large number of experimental
facilities which can provide precision measurements
of many physical quantities related to the underlying spin structure
of hadrons. Among the various topics the strange content
of the proton is one of the most attractive due to its
close connections to the proton spin problem \cite{Bro88} and
to the quark-antiquark asymmetry of the nucleon sea \cite{Bro96}.
Though there have been
many progresses and achievements, 
both theoretically and experimentally,
our current knowledge of the strange quark content
of the proton is still very poor, since one is still unclear, 
whether or not \cite{Bro88,Bro96}, strange quarks are highly polarized
inside the proton, and   
it is even more obscure, whether or not \cite{Bro96,Bro97}, 
the strange quark-antiquark 
distributions are symmetric.
Thus a precision measurement of the strange-antistrange 
polarizations of the proton is one of the most challenging 
and significant tasks for 
hadron quark 
structure.  

 From another point of view, the quark distribution of a quark inside
a hadron is related by crossing to the fragmentation function
of the same flavor quark to the same hadron, by a simple reciprocity
relation \cite{GLR}
\begin{equation}
q_{h}(x) \propto D_q^h(z),
\end{equation}
where $z=2 p \cdot q/Q^2$ is the momentum fraction of the produced
hadron from the quark jet in the fragmentation process, 
and $x=Q^2/2 p \cdot
q$ is the Bjorken scaling variable corresponding to the momentum
fraction of the quark from the hadron in the deep inelastic
scattering (DIS) process.
Although such an approximate relation may be only valid
at a specific scale $Q^2$, which deserves
further studies, 
it could provide reasonable connection between 
different physical quantities and lead to different
predictions about the fragmentations based on our understanding
of the quark structure of a hadron \cite{Bro97}. 
Thus measurements of quark fragmentations into a hadron
can also provide some new insights into the quark structure of the hadron.
Among the various produced hadrons, $\Lambda$ hyperon 
is most suitable to study the polarized fragmentation due
to its self-analyzing property owing to 
the characteristic decay mode $\Lambda \to p \pi^-$ 
with a large branching ratio of 64\%. The technology to disentangle
between $\Lambda$ and $\bar{\Lambda}$,
which relies on magnetic spectrometers,
is also mature
for DIS processes \cite{Lbar}. In fact there are already some  
available data
on $\Lambda$ and $\bar{\Lambda}$ productions in different
DIS processes and more to come in the near future
\cite{Data,Nomad,Hermes,Compas}.
Therefore it is meaningful and urgent to exploit a
systematic way to extract the various quark contributions to $\Lambda$ 
fragmentation functions.

There have been many proposals concerning the measurements    
of the $\Lambda$ fragmentations functions in different processes, 
for different physical goals [2,10-21], 
and in this letter we will focus
our attention on the longitudinally polarized case. One promising method
to obtain a complete set of polarized fragmentation functions
for different quark flavors
is based on the measurement of the helicity asymmetry 
for semi-inclusive production of $\Lambda$ hyperons in
$e^+e^-$ annihilation on the $Z^0$ resonance \cite{Bur93}, 
but the existing data
can only provide a poor constraint for different scenarios
\cite{Flo98b}.
Measurements of the light-flavor quark fragmentations into $\Lambda$
have been also suggested from polarized electron
DIS process \cite{Jaf96} and neutrino DIS process
\cite{Kot97}, based on the $u$-quark dominance
assumption. 
There is also a recent interesting suggestion to determine
the polarized fragmentation functions by measuring the helicity transfer
asymmetry in the process
$p \overrightarrow{p} \to \overrightarrow{\Lambda} X$ 
\cite{Flo98}. From its dependence on the
rapidity of the $\Lambda$, it is possible to discriminate
between various parametrizations. 
In this letter we will show
that the neutrino and antineutrino
deep inelastic scattering processes of
unpolarized and polarized $\Lambda$
and $\bar{\Lambda}$ productions can provide a
clean separation of unpolarized and polarized 
fragmentation functions 
of a light-flavor quark into a $\Lambda$, 
for both quarks and antiquarks, 
and also for strange quarks. 
Combining with
polarized electron beam DIS
processes of unpolarized and 
polarized $\Lambda$ and $\bar{\Lambda}$ productions, 
one can systematically measure or check the various 
flavor and spin dependent fragmentation functions.  
Thus, in addition to the 
known process 
$e^+e^- \to \overrightarrow{\Lambda} X $ \cite{Bur93}, 
we have a different method,  
to measure a complete
set of quark to $\Lambda$ unpolarized and polarized
fragmentation functions for different quark flavors
by the systematic exploitation of unpolarized and polarized 
$\Lambda$ and $\bar{\Lambda}$
productions in neutrino, antineutrino and polarized
electron DIS processes.

Our considerations rely on the fact 
that neutrinos (antineutrinos) can be regarded
as a purely polarized lepton beam, 
due to the fact that neutrinos are
left-handed (antineutrinos are right-handed), 
therefore they only
interact with quarks of specific helicities and flavors.
For example, a neutrino (antineutrino) can only interact with the
$d$, $\bar{u}$, and $s$ ($u$, $\bar{d}$ and $\bar{s}$) 
light-flavor quarks with left-handed 
quarks 
and with right-handed 
antiquarks 
of a hadronic target, regardless 
this target is polarized or not,
and the scattered quarks will keep the same helicities of
their parent quarks before the collision \cite{Bigi}. 
Thus the scattering of a neutrino (antineutrino) 
beam on a hadronic target, 
provides a {\it source of polarized quarks with specific flavor
structure}, and this particular property makes 
neutrino (antineutrino) DIS process, an ideal laboratory
to study the flavor-dependence quark fragmentation  
into hadrons in the current fragmentation region, 
especially in the polarized case. 

 From the charged current quark transitions, 
for neutrino induced reactions
\begin{equation}
\begin{array}{clcr}
\nu~ d \to \mu^- u; ~~~~~~~~~~ \nu~ d \to \mu^- c; \\
\nu~ \bar{u} \to \mu^- \bar{d}; 
~~~~~~~~~~ \nu~ \bar{u} \to \mu^- \bar{s};\\
\nu~ s \to \mu^- c; ~~~~~~~~~~ \nu~ s \to \mu^- u,
\end{array}
\end{equation}
and for antineutrino induced reactions
\begin{equation}
\begin{array}{clcr}
\bar{\nu}~ u \to \mu^+ d; ~~~~~~~~~~ \bar{\nu}~ u \to \mu^+ s; \\
\bar{\nu}~ \bar{d} \to \mu^+ \bar{u}; 
~~~~~~~~~~ \bar{\nu}~ \bar{d} \to \mu^+ \bar{c};\\
\bar{\nu}~ \bar{s} \to \mu^+ \bar{c}; 
~~~~~~~~~~ \bar{\nu} ~\bar{s} \to \mu^+ \bar{u},
\end{array}
\end{equation} 
the expressions for the $\Lambda$ and $\bar{\Lambda}$ longitudinal
polarizations in the beam direction are,
for $\Lambda$ and $\bar \Lambda$
produced in the current fragmentation,
\begin{equation}
P_{\nu}^{\Lambda}(x,y,z)=
-\frac{d(x)\Delta D_u^{\Lambda}(z)-(1-y)^2 \bar{u}(x)\Delta D
_{\bar{d}}^{\Lambda}(z)}
{d(x) D_u^{\Lambda}(z)+ (1-y)^2 \bar{u}(x) D
_{\bar{d}}^{\Lambda}(z)}
~~~~{\it for}~~~~{\nu N \to \mu^- \overrightarrow{\Lambda} X}; 
\label{nL}
\end{equation} 
\begin{equation}
P_{\bar{\nu}}^{\Lambda}(x,y,z)=
-
\frac{(1-y)^2 u(x)\Delta D_d^{\Lambda}(z)-\bar{d}(x)\Delta D
_{\bar{u}}^{\Lambda}(z)}
{(1-y)^2 u(x) D_d^{\Lambda}(z)+ \bar{d}(x) D
_{\bar{u}}^{\Lambda}(z)}
~~~~{\it for}~~~~{\bar{\nu} N \to \mu^+ \overrightarrow{\Lambda} X}; 
\label{nbL}
\end{equation} 
\begin{equation}
P_{\nu}^{\bar{\Lambda}}(x,y,z)=
-\frac{d(x)\Delta D_u^{\bar{\Lambda}}(z)-(1-y)^2\bar{u}(x)\Delta D
_{\bar{d}}^{\bar{\Lambda}}(z)}
{d(x) D_u^{\bar{\Lambda}}(z)+ (1-y)^2 \bar{u}(x) D
_{\bar{d}}^{\bar{\Lambda}}(z)}
~~~~{\it for}~~~~{\nu N \to \mu^- \overrightarrow{\bar{\Lambda}} X}; 
\label{nLb}
\end{equation} 
\begin{equation}
P_{\bar{\nu}}^{\bar{\Lambda}}(x,y,z)=
-
\frac{(1-y)^2u(x)\Delta D_d^{\bar{\Lambda}}(z)-\bar{d}(x)\Delta D
_{\bar{u}}^{\bar{\Lambda}}(z)}
{(1-y)^2u(x) D_d^{\bar{\Lambda}}(z)+ \bar{d}(x) D
_{\bar{u}}^{\bar{\Lambda}}(z)}
~~~~{\it for}~~~~{\bar{\nu} N \to \mu^+ \overrightarrow{\bar{\Lambda}}
X}. 
\label{nbLb}
\end{equation} 
Here we have neglected the Cabibbo suppressed
processes and the small strange quark distributions 
inside the hadronic target,
$\Delta D_q^h(z)=D_{q\uparrow}^{h\uparrow}(z)-
D_{q\uparrow}^{h\downarrow}(z)$ denotes the polarized fragmentation
function, $D_{q\uparrow}^{h\uparrow}(z)$ 
($D_{q\uparrow}^{h\downarrow}(z)$) being the
probability for finding a hadron with positive (negative)
helicity in a quark $q$ with positive helicity, and $y=\nu/E$
is the energy fraction of the incident neutrino 
carried by the charged intermediate boson $W^{\pm}$, 
in the laboratory frame. 
For any value of $x$ and $z$, we see that the measurement
of these four quantities in the region $y \simeq 1$,
leads directly to the two 
fragmentation asymmetries
$\Delta D_{q}^{\Lambda}(z)/ D_{q}^{\Lambda}(z)$, where
$q=u$ for processes (\ref{nL}) and (\ref{nbLb})
and $q=\bar{u}$ for processes (\ref{nbL}) and (\ref{nLb}).
Of course here
we have 
applied 
matter and antimatter
symmetry, i.e. $D_{q,\bar{q}}^{\Lambda}(z)
=D_{\bar{q},q}^{\bar{\Lambda}}(z)$ and
similarly for $\Delta D_{q,\bar{q}}^{\Lambda}(z)$. We can also assume
the $u$ and $d$ symmetry for the
fragmentation functions $D_{u}^{\Lambda,\bar{\Lambda}}(z)=
D_{d}^{\Lambda,\bar{\Lambda}}(z)$ 
and $D_{\bar{u}}^{\Lambda,\bar{\Lambda}}(z)=
D_{\bar{d}}^{\Lambda,\bar{\Lambda}}(z)$,
from the symmetry of $u$ and $d$ inside $\Lambda$
and $\bar{\Lambda}$. Hence we have, for
the unpolarized fragmentation functions 
\begin{equation}
\begin{array}{clcr}
D_q^{\Lambda}(z)=D_u^{\Lambda}(z)=D_d^{\Lambda}(z)
=D_{\bar{u}}^{\bar{\Lambda}}(z)=D_{\bar{d}}^{\bar{\Lambda}}(z);\\
D_{\bar{q}}^{\Lambda}(z)
=D_{\bar{u}}^{\Lambda}(z)=D_{\bar{d}}^{\Lambda}(z)
=D_u^{\bar{\Lambda}}(z)=D_d^{\bar{\Lambda}}(z),
\label{uff}
\end{array}
\end{equation}
and for the polarized fragmentation functions
\begin{equation}
\begin{array}{clcr}
\Delta D_q^{\Lambda}(z)
=\Delta D_u^{\Lambda}(z)=\Delta D_d^{\Lambda}(z)
=\Delta D_{\bar{u}}^{\bar{\Lambda}}(z)
=\Delta D_{\bar{d}}^{\bar{\Lambda}}(z);\\
\Delta D_{\bar{q}}^{\Lambda}(z)
=\Delta D_{\bar{u}}^{\Lambda}(z)=\Delta D_{\bar{d}}^{\Lambda}(z)
=\Delta D_u^{\bar{\Lambda}}(z)=\Delta D_d^{\bar{\Lambda}}(z).
\label{pff}
\end{array}
\end{equation}
Eqs.(\ref{nL}) to (\ref{nbLb}) thus reduce to
equations of the four fragmentation functions
$D_q^{\Lambda}(z)$, $D_{\bar{q}}^{\Lambda}(z)$,
$\Delta D_{q}^{\Lambda}(z)$, and $\Delta D_{\bar{q}}^{\Lambda}(z)$.
The cross sections
of the corresponding unpolarized $\Lambda$ and $\bar{\Lambda}$ 
productions can be written
\begin{equation}
\frac{1}{\sigma_{\mathrm tot}}\frac{{\mathrm d} \sigma}
{ {\mathrm d} z}
=\frac{\sum_i \left[ a_i q_i(x)+ b_i \bar{q}_i(x)\right] 
 D_i^{\Lambda,\bar{\Lambda}}(z)}
{\sum_i \left[ a_i q_i(x) +b_i \bar{q}_i(x) \right]},
\end{equation}
where $i$ implies all quark (antiquark) flavors
involved in the corresponding process ($i=d,\bar{u}$ for neutrino
beams and $i=u,\bar{d}$ for antineutrino beams, neglecting
the Cabibbo mixings and the strange distributions of the target
in the formula)
and $a_i$ and $b_i$ are two factors with $a_i=1$ and $b_i=0$
($a_i=1/3$ and $b_i=0$)
for relevant quarks and $a_i=0$ and $b_i=1/3$ 
($a_i=0$ and $b_i=1$) for antiquarks
for neutrino (antineutrino)
induced processes \footnote{We can replace in the
numerator
the factors $1/3$ by $(1-y)^2$ for $y$-dependent
cross sections, and the factors $1$ by $\int_{y_{l}}^{y_{h}} d y$
and the factors $1/3$ by $\int_{y_{l}}^{y_{h}} d y~(1-y)^2$
to take into the experimental cuts for $y_l < y < y_h$.}. 

In principle, Eqs.~(\ref{nL})-(\ref{nbLb}) are sufficient 
to determine
all four independent fragmentation functions
$D_q^{\Lambda}(z)$, $D_{\bar{q}}^{\Lambda}(z)$,
$\Delta D_{q}^{\Lambda}(z)$, and $\Delta D_{\bar{q}}^{\Lambda}(z)$,
at specified $x$ and $y$, or integrating both
the numerators and denominators over $y$ and/or $x$
within the experimental cuts.
However, in practice we must be careful, since
the contribution from Cabibbo suppressed process
of $u \to s \to \Lambda$ 
($\bar {u} \to \bar{s} \to \bar{\Lambda}$) 
is by no means negligible 
in the process (\ref{nbL}) $\bar{\nu} N \to \mu^+\Lambda X$ 
((\ref{nLb}) $\nu N \to \mu^- \bar{\Lambda} X $).
By measuring both the unpolarized and polarized
$\Lambda$ and $\bar \Lambda$ productions, using 
the relatively clean processes (\ref{nL}) and (\ref{nbLb}), it seems
enough to determine the four independent fragmentation functions.
Nevertheless, as already mentioned above,
by varying $y$ and/or $x$, 
if data allow it, one may make a
clean flavor separation 
(even without 
the assumption of $u$ and $d$ flavor
symmetry) 
by only the polarized processes (\ref{nL}) and (\ref{nbLb}). 
The antiquark contribution of the target 
can be only safely neglected  
at large $y$ for process (\ref{nL}), not for processes
(\ref{nbL}) and (\ref{nbLb}) where it dominates.
 
There are many possible ways to obtain
the strange fragmentation functions
from the neutrino,
antineutrino and polarized electron DIS processes. 
The first way
is to use processes (\ref{nL}) and (\ref{nbLb}),
$y$ and/or $x$ dependent (or varying the cuts for $x$
and $y$), or
polarized combined with unpolarized ones within
all the cuts, 
to extract the four
light-flavor quark fragmentation functions.
Then by substituting these four quantities 
in polarized 
(\ref{nbL}) and (\ref{nLb}) (or a $y$ and/or $x$ dependent single
process),
one can determine the strange fragmentation functions
$D_s^{\Lambda}$ and $\Delta D_s^{\Lambda}$ \footnote{
The fragmentation process $\bar{s} \to \Lambda$ ($s \to
\bar{\Lambda}$) is expected to be much smaller compared
to $s \to \Lambda$ ($\bar{s} \to \bar{\Lambda}$),
since the former is from the sea $s\bar{s}$ pairs
in $\Lambda$, whereas the latter is from
the dominant valence configuration
in which $s$ provides the spin of $\Lambda$.}.
Besides, with 
unpolarized
$\Lambda$ production of process (\ref{nbL}) 
one can also, in principle, 
extract the strange fragmentation function 
$D_s^{\Lambda}(z)$ 
provided $D_q^{\Lambda}(z)$ and 
$D_{\bar{q}}^{\Lambda}(z)$ are known and then check 
this quantity
by using (\ref{nLb}) \footnote{ One can also use 
the unpolarized processes 
(\ref{nL}) to (\ref{nbLb}) 
to determine $D_q^{\Lambda}$,
$D_{\bar{q}}^{\Lambda}$, $D_s^{\Lambda}$,
and $D_{\bar{s}}^{\Lambda}$,
if one thinks that
$D_{\bar{s}}^{\Lambda}$ is not negligible, and then by extending
the same analysis to the polarized processes to measure
$\Delta D_q^{\Lambda}$,
$\Delta D_{\bar{q}}^{\Lambda}$, $\Delta D_s^{\Lambda}$,
and $\Delta D_{\bar{s}}^{\Lambda}$.}.
Then by extending the same analysis
to the polarized cases one can also first measure
$\Delta D_s^{\Lambda}(z)$ in process (\ref{nbL}) and then check
this quantity by process (\ref{nLb}). 
Thus the measurements
of unpolarized and polarized  $\Lambda$ and 
$\bar{\Lambda}$ productions in neutrino and antineutrino
can in principle provide a full determination of the 
flavor-dependent unpolarized and polarized fragmentation 
functions. 

Another way is to combine 
the polarized
$\Lambda$ and $\bar{\Lambda}$ productions
in neutrino and antineutrino
DIS processes with those of polarized electron
DIS processes \cite{Jaf96} and determine
the four light-flavor fragmentation functions and 
two strange fragmentation
functions from six independent equations of
$\Lambda$ ($\bar{\Lambda}$) polarizations of
the six processes \footnote{New high statistics data
are expected soon from several experiments \cite{Hermes,Compas}.}. 
This method has the advantage that one does not need to take care
of the magnitudes or the $y$ and/or $x$ dependences of 
the $\Lambda$ and $\bar{\Lambda}$ 
productions but only their polarizations.
Needless to say that the strange (antistrange)
quark distributions of the hadronic target in the polarized
electron DIS processes must be taken into
account due to the large fragmentation process 
$s \to s \to \Lambda$ 
($\bar{s} \to \bar{s} \to \bar{\Lambda}$), as
we have mentioned. It is interesting to note
that the use of different nuclear targets,
with different isospin properties,
can also help us to gain additional information
on the flavor-dependence for the fragmentation functions.

Clearly, the above analysis can be also extended to the
productions of other hadrons, such as $\Sigma$,
$\Xi$  et al,    
or even to heavier flavor hadrons such as $\Lambda_c$. 
Though such studies might be tedious theoretically 
and difficult experimentally, they
are meaningful not only for their own purpose but, also
for the physics purpose of this paper. 
The reason is that the final $\Lambda$ 
and $\bar{\Lambda}$ hyperons might come from decays of
these hadrons which generate a background to be studied and
removed for a precise understanding of
the quark to $\Lambda$ fragmentation.
Notice that some progress has been already
achieved along this direction \cite{Bigi,Gus93,Kot97,Bor98},
also on the QCD evolution 
of the fragmentation functions \cite{Flo98b}
and on higher order corrections \cite{Ans96}.
It should be pointed out that the presented
results in Eqs.(\ref{nL})-(\ref{nbLb}) hold only in leading order.
In next-to-leading order the $x/z$ factorization does not hold
any longer, which complicates the analysis considerably.
Also processes induced by gluons and the fragmentation
of gluons into $\Lambda$ and $\bar \Lambda$ become relevant.
All these effects, including also the contributions
from  Cabibbo mixings, will have to be further taken into account with 
increasing statistical accuracy of the data
\footnote{To our knowledge, the highest statistics
data on $\Lambda$ and $\bar \Lambda$ production separately
comes from the
NOMAD neutrino beam experiment \cite{Nomad}.}. 

It is known that the original proposal \cite{Lu95} for measuring
the strange quark polarization of a proton $\Delta s(x)$ from
the semi-inclusive $\Lambda$ polarization of
unpolarized electron beam on polarized proton
target DIS process, suffers from the contributions  
of $u$ and $d$ quark fragmentations due to the non-zero
$\Delta D_q^{\Lambda}(z)$ \cite{Jaf96,Kot97,Flo98b}. 
With a better knowledge of the flavor and spin dependent
quark to $\Lambda$ fragmentation functions, one can try to
remove the background from contributions of 
the polarized $u$ and $d$ quarks 
of the proton to the polarized $\Lambda$,
to allow the measurement of  $\Delta s(x)$.
As also mentioned in Ref.\cite{Bro96}, a complementary
measurement of $\bar{\Lambda}$ polarization in the same
process can also help to pin down the antistrange  
quark polarization inside the
proton.\footnote{Remember that neutrino (antineutrino)
DIS process cannot be done on a polarized
target and therefore cannot provide any
information concerning the spin structure of the nucleon target.}. 
Such measurements can provide crucial tests of different
predictions concerning the spin structure of hadrons
\cite{Bro88,Bro96}
and the quark-antiquark asymmetry of the nucleon
sea \cite{Bro96,Bro97}.
Besides, the $u$ and $d$ polarizations inside a $\Lambda$
are closely related to the physical mechanism for
the $s$ polarization inside a proton. Measurement
of $\Delta D_q^{\Lambda}$ is also helpful to understand
the mechanism for the $s$ polarization of the proton. 

We need to point out that the measurement of $\Delta
D_{\bar{q}}^{\Lambda}$ is also related to the quark-antiquark
asymmetry discussed in Refs.\cite{Bro96,Bro97}. From the
SU(3) symmetry argument of Burkardt-Jaffe \cite{Bur93} we know that
the $u$ and $d$ quarks inside a $\Lambda$ 
should be negatively polarized. If
the valence $u$ and $d$ in a $\Lambda$ are unpolarized, one may
simply expect the $u$ and $d$ polarizations coming from the sea, 
with the same polarizations for quarks and antiquarks, 
for the $u$ and $d$ flavors,
and therefore $\Delta
D_{\bar{q}}^{\Lambda} \neq 0$. However, this might not be true
in general and in the baryon-meson
fluctuation model of the intrinsic sea quarks 
of a hadron \cite{Bro96},
the intrinsic $u\bar{u}$ and $d\bar{d}$ pairs inside a
$\Lambda$ are mainly from the configurations $\Lambda(udsu\bar{u})=
p(uud)K^-(s\bar{u})$ and $\Lambda(uds d \bar{d})=
n(udd)K(s\bar{d})$. 
 From this picture the $u$ and $d$ quarks inside
a $\Lambda$ are negatively polarized, whereas $\bar{u}$
and $\bar{d}$ should be unpolarized or slightly polarized
from higher fluctuations. In fact, the predictions
\cite{Bro96,Che95} of
a small or zero polarization of the sea antiquarks
in the proton, are supported by the Spin Muon Collaboration
measurement of the $u$ and $d$ antiquark helicity distributions
from semi-inclusive DIS process \cite{NSMCN}. 
Therefore
the future measurement of $\Delta D_{\bar{q}}^{\Lambda}$ can provide
another test of different predictions concerning the hadron spin structure. 
For example, the $\Lambda$ and $\bar \Lambda$ polarizations 
for processes Eq.(\ref{nL}) and Eq.(\ref{nLb})
at $y \simeq 1$	are predicted to be 
$P_{\nu}^{\Lambda}=0.14 \pm 0.04$ and  
$P_{\nu}^{\bar \Lambda}=0.06 \pm 0.02$ for case I and
$P_{\nu}^{\Lambda}=0.09 \pm 0.04$ and  
$P_{\nu}^{\bar \Lambda}=0.12 \pm 0.04$ for case II
in Ref.\cite{Jaf96},
whereas they are $P_{\nu}^{\Lambda} \approx 0.02$ 
and $P_{\nu}^{\bar \Lambda}=0$ in the baryon-meson fluctuation model
assuming an intrinsic $u \bar u$ sea fluctuation probability
around 10\% and $\Delta u_{N}+\Delta d_{N} \approx 0.5$ \cite{Bro96}, 
as shown in Fig.\ref{msof1}.

\begin{figure}[htb]
\begin{center}
\leavevmode {\epsfysize=5.2cm \epsffile{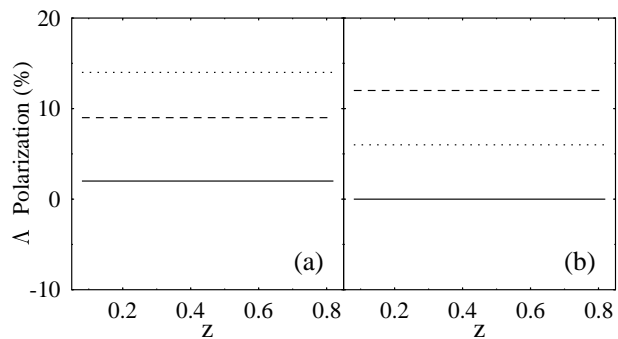}} 
\end{center}            
\caption[*]{\baselineskip 13pt 
The $\Lambda$ and $\bar \Lambda$ polarization for processes
Eq.(\ref{nL}) (a) and Eq.(\ref{nLb}) (b)
at $y \simeq 1$. The dotted and dashed lines are the predictions
for cases I and II in Ref.\cite{Jaf96} with a simple ansatz
$\Delta D_{\bar{q}}^{\Lambda}(z)
=C_{\bar{q}} D_{\bar{q}}^{\Lambda}(z)$ of a constant
coefficient $C_{\bar{q}}$. The solid lines 
are the prediction of the baryon-meson
model \cite{Bro96}. 	
}
\label{msof1}
\end{figure}

In summary, in this letter we showed that 
a complete investigation of $\Lambda$ and $\bar{\Lambda}$ hyperon
productions in neutrino and antineutrino DIS processes
can provide an ideal laboratory for a systematic study
of the flavor and spin dependence of the quark fragmentations.
The flavor and spin separation we propose, 
is based on the particular property that 
the scattering of neutrinos (antineutrinos) 
on a hadronic target
provides a source of polarized quarks with specific flavor
structure. A full determination of unpolarized
and polarized quark to $\Lambda$ fragmentation functions
is also helpful, to measure the strange and antistrange
polarizations of the proton and to test ideas related to the
spin structure of the nucleon and to the quark-antiquark
asymmetry of the nucleon sea.   

{\bf Acknowledgments: } 
We would like to thank 
S.J.~Brodsky,
A.~Kotzinian,
and 
F.~Montanet for helpful discussions 
and C.Joseph for useful informations about NOMAD.
This work is 
partially supported by National Natural 
Science Foundation of China 
under Grant No.~19605006.

\end{document}